\documentclass[runningheads]{llncs}
\usepackage{booktabs}
\usepackage{multirow}
\usepackage{amssymb}
\usepackage[T1]{fontenc}
\usepackage{hyperref}

\usepackage{graphicx}
\begin{document}
\title{VoxDub: A Bilingual Lip-Synchronized Audio-Visual Dataset for Automatic Video Dubbing\thanks{This work was supported in part by the National Natural Science Foundation of China under Grants 62276220 and 62371407 and the Innovation of Policing Science and Technology, Fujian province (Grant number: 2024Y0068).}}
\titlerunning{VoxDub: A Bilingual Dataset for Automatic Video Dubbing}

\author{Kaidi Wang\inst{1} \and
Wenhao Guan\inst{2} \and
Peijie Chen\inst{1} \and
Weijie Wu\inst{1} \and
Xuetao Lin\inst{1} \and
Lin Li\inst{2,\dagger} \and
Qingyang Hong\inst{1,\dagger}
}

\authorrunning{K. Wang et al.}

\institute{
School of Informatics, Xiamen University, China\\
\and
School of Electronic Science and Engineering, Xiamen University, China\\[0.5ex]
\email{qyhong@xmu.edu.cn}
}
%
\begingroup
\renewcommand{\thefootnote}{\ensuremath{\dagger}}
\footnotetext[0]{Corresponding authors}
\endgroup
\maketitle              
\begin{abstract}
Automatic video dubbing (AVD) requires temporally aligned speech, transcripts, and video, yet existing corpora vary in audio quality, transcription reliability, and speaking-face visibility. We introduce VoxDub, a bilingual lip-synchronized audio-visual dataset containing 766K clips and approximately 1.58K hours of in-the-wild Chinese and English recordings, covering varied speakers, scenes, speaking styles, and linguistic content. Its curation pipeline combines speech-centered segmentation with dual-ASR transcription checking, speech quality filtering, face visibility checks, and lip-speech synchronization verification. Text-to-speech (TTS) experiments show that training on VoxDub yields lower word error rates and higher speaker similarity than training on the compared audio-visual corpora, LRS3 and CN-CVS. AVD experiments show broadly comparable speech quality and subjective synchronization. These results support the utility of VoxDub for TTS and AVD training. More details are available on the \href{https://voxdub-dataset.github.io/VoxDub/}{project website}.

\end{abstract}
\section{Introduction}

Automatic video dubbing (AVD)~\cite{hu2021neural} aims to generate natural speech that matches the linguistic content, speaking style, duration, and lip movements of a target video. Compared with conventional text-to-speech (TTS), AVD additionally requires temporal alignment between generated speech and visible facial motion. Training data must therefore provide not only usable speech--text pairs but also corresponding visual information with consistent lip-speech alignment.

Existing resources meet different parts of this requirement. Speech and audio-visual datasets~\cite{chung2018voxceleb2,zhang2021flow} have supported a range of modeling tasks, while TTS-oriented corpora such as Emilia~\cite{he2024emilia} provide speech and transcriptions without visual supervision. Audio-visual corpora supply paired speech and video, but differ in their treatment of audio quality, transcription reliability, language and domain coverage, and continuous face visibility. Recent work on joint audio-motion generation and video dubbing~\cite{kwon2025jam,sung2025voicecraft} further motivates the use of multimodal training data. For the setting considered here, the data curation problem is to retain usable speech segments while checking the quality and consistency of the audio, text, and visible speaking face across both Chinese and English recordings.

We address this problem with \textbf{VoxDub}, a bilingual corpus of lip-synchronized video--speech pairs for AVD. It contains more than 766K clips and approximately 1.58K hours of in-the-wild recordings spanning varied speakers, scenes, speaking styles, and linguistic content. The pipeline first segments recordings around speech activity, then filters candidates using transcription agreement, speech quality, face visibility, and lip-speech synchronization. This ordering prioritizes speech segments while applying visual and synchronization checks to the corresponding video intervals.

We assess the resulting corpus through dataset statistics and two complementary downstream tasks. TTS experiments evaluate the utility of its speech--text pairs, whereas AVD experiments evaluate its use for adapting a pretrained speech generator to video-conditioned dubbing. The main contributions are as follows:
\begin{itemize}
    \item We construct VoxDub, a bilingual audio-visual dataset of more than 766K lip-synchronized clips for automatic video dubbing.
    \item We develop a curation pipeline that combines speech-centered segmentation with checks on audio quality, transcription agreement, face visibility, and lip-speech synchronization.
    \item We characterize the corpus and evaluate it on TTS and AVD, reporting improved TTS performance relative to the compared audio-visual corpora and competitive AVD performance under the evaluated settings.
\end{itemize}

\section{Related Work}

\subsection{Text-to-Speech Datasets}

Speech corpora provide the audio and text supervision required for TTS. LibriTTS~\cite{Zen2019LibriTTSAC}, derived from LibriSpeech~\cite{panayotov2015librispeech}, contains 585 hours of English multi-speaker read speech with normalized text. Emilia~\cite{he2024emilia} extends this setting to large-scale multilingual in-the-wild speech, using source separation, voice activity detection (VAD), automatic speech recognition (ASR), and filtering to construct speech--text pairs. These corpora are relevant to the speech quality and transcription requirements of AVD, but their audio-only format does not provide visual supervision for learning the correspondence between speech timing and facial motion.

\subsection{Audio-Visual Datasets}

Audio-visual speech datasets serve tasks including lip reading, talking-face generation, speaker recognition, and audio-visual speech modeling. LRS3~\cite{afouras2018lrs3}, CMLR~\cite{zhao2019cascade}, and CN-CVS~\cite{10095796} provide paired speech and video for visual speech recognition and related tasks. Table~\ref{tab:dataset_comparison} places these resources alongside speech-only corpora and datasets used for talking-face generation and dubbing. Their different intended uses lead to different choices of segmentation, annotation, language coverage, and audio-visual filtering; they should therefore not be treated as interchangeable sources of AVD training data.

VoxDub focuses on speech-centered clips with explicit checks on transcription agreement, audio quality, face visibility, and lip-speech synchronization. Its contribution is the combination of these checks in a Chinese--English corpus, rather than synchronization filtering alone. In Table~\ref{tab:dataset_comparison}, ``Seg. Modality'' denotes the primary modality used to segment long recordings, and ``Full Lip-Speech Align.'' summarizes the treatment of lip-speech correspondence across a clip. ``Explicit'' denotes an explicit check or filtering step, ``Assumed'' denotes synchronization inherited from the source recordings without explicit filtering, and ``Partial'' denotes alignment constraints that do not cover the entire clip. These categories describe curation procedures rather than a common quantitative measure of alignment quality.

\begin{table*}[t]
\centering
\caption{Comparison with representative speech and audio-visual datasets.}
\label{tab:dataset_comparison}
\resizebox{\textwidth}{!}{
\begin{tabular}{lccccccc}
\hline
Dataset & Language & Scale & Transcript & Visual & Data Source & Seg. Modality & Full Lip-Speech Align. \\
\hline
LibriTTS~\cite{Zen2019LibriTTSAC} & EN & 585 h & \checkmark & $\times$ & Audiobook & Audio & -- \\
Emilia~\cite{he2024emilia} & Multi & 101K+ h & \checkmark & $\times$ & In-the-wild & Audio & -- \\
MEAD~\cite{kaisiyuan2020mead} & EN & 40+ h & \checkmark & \checkmark & Studio recording & -- & -- \\
HDTF~\cite{zhang2021flow} & EN & 15.8 h & $\times$ & \checkmark & In-the-wild & Visual & Assumed \\
CMLR~\cite{zhao2019cascade} & ZH & 61+ h & \checkmark & \checkmark & In-the-wild & Audio & Assumed \\
LRS3~\cite{afouras2018lrs3} & EN & 400+ h & \checkmark & \checkmark & In-the-wild & Visual & Explicit \\
CN-CVS~\cite{10095796} & ZH & 300+ h & \checkmark & \checkmark & In-the-wild & Visual & Explicit \\
VoxCeleb2~\cite{chung2018voxceleb2} & Multi & 2K+ h & $\times$ & \checkmark & In-the-wild & Visual & Explicit \\
CelebV-Dub~\cite{sung2025voicecraft} & EN & 86 h & \checkmark & \checkmark & In-the-wild & Visual & Explicit \\
CineDub-CN~\cite{liu2026funcineforge} & ZH & 4.7K h & \checkmark & \checkmark & In-the-wild & Audio & Partial \\
\textbf{VoxDub} & \textbf{ZH/EN} & \textbf{1.58K h} & \checkmark & \checkmark & \textbf{In-the-wild} & \textbf{Audio} & \textbf{Explicit} \\
\hline
\end{tabular}
}
\end{table*}

\section{VoxDub Dataset}

\subsection{Construction Pipeline}
\begin{figure}[t]
\centering
\includegraphics[width=\linewidth]{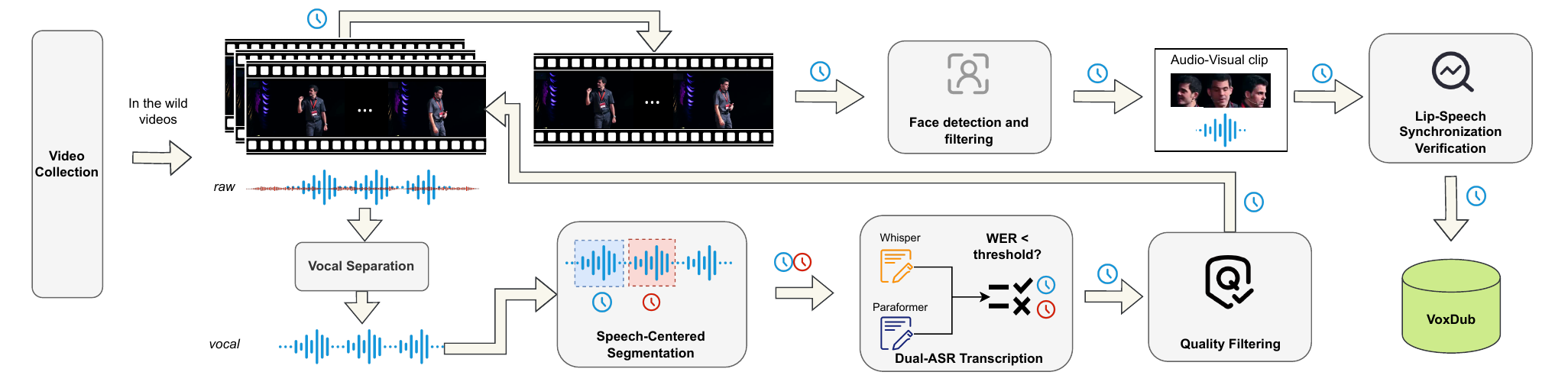}
\caption{Overview of the VoxDub dataset construction pipeline.}
\label{fig:pipeline}
\end{figure}

The VoxDub construction pipeline is illustrated in Fig.~\ref{fig:pipeline}. Audio processing first produces candidate speech segments and transcriptions. Visual filtering then checks face visibility and lip-speech correspondence within the matching video intervals. The stages are described below.

\subsubsection{Video Collection}

We collect videos from online streaming platforms using manually selected topic keywords covering talks, interviews, films, news, vlogs, and daily conversations. For each candidate, we record the video ID and metadata and download the highest-quality available video and audio streams. We standardize the recordings to 25 frames per second (FPS) and a 24 kHz audio sampling rate before further processing.

\subsubsection{Vocal Separation}
In-the-wild recordings can contain background music, environmental noise, and other non-speech sounds that interfere with transcription and speech generation. Following Emilia~\cite{he2024emilia}, we use Ultimate Vocal Remover (UVR)\footnote{\url{https://github.com/Anjok07/ultimatevocalremovergui}} to separate each audio track into vocal and non-vocal components. Only the vocal track is retained for subsequent processing.

\subsubsection{Speech-Centered Segmentation}
VoxDub uses speech activity, rather than visual tracks, as the primary basis for segmentation. This choice is intended to reduce speech truncation and retain segments suitable for transcription and speech generation. We apply VAD\footnote{\url{https://github.com/snakers4/silero-vad}} to the separated vocal track to obtain timestamps for speech-active regions. Regions separated by short pauses are merged, and overly short segments are removed to reduce over-segmentation. We then extract the corresponding video intervals using the same timestamps, preserving the temporal correspondence between each audio segment and its video.

\subsubsection{Dual-ASR Transcription}
We use agreement between two ASR systems as a transcription filtering criterion. WhisperX~\cite{bain2022whisperx}, using Whisper-large-v3~\cite{pmlr-v202-radford23a}, and Paraformer~\cite{Gao2022ParaformerFA} are applied to all candidate speech segments. WhisperX provides transcription and language identification; only segments identified as Chinese or English are retained. For each retained segment, we compute the word error rate (WER) between the two ASR hypotheses and discard candidates whose WER exceeds a predefined threshold. The final annotation is the Whisper transcription for English and the Paraformer transcription for Chinese.

\subsubsection{Speech Quality Filtering}

We filter the candidate audio to reduce noise, distortion, and other acoustic artifacts. Specifically, we use the overall quality (OVRL) score from DNSMOS P.835~\cite{reddy2022dnsmos} and discard clips with scores below 2.8. To reduce speaker inconsistency within clips, we also apply a multi-speaker detection model based on the CED architecture~\cite{dinkel2023ced}. Segments detected as containing multiple speakers or speaker changes are removed.

\subsubsection{Face Detection and Filtering}

We apply frame-level face detection~\cite{deng2020retinaface} and retain only clips in which exactly one face is detected in every frame. This criterion favors continuous face visibility and reduces ambiguity associated with missing or multiple visible faces. We also compare face bounding boxes in adjacent frames and discard clips with abrupt location changes, which may indicate a face switch, shot transition, or unstable detection. These checks select clips with more stable face detections.

\subsubsection{Lip-Speech Synchronization Verification}
Continuous face detections do not by themselves establish that the visible face corresponds to the speech. We therefore use SyncNet\footnote{\url{https://huggingface.co/ByteDance/LatentSync-1.6/tree/main/auxiliary}}~\cite{Chung16a} to estimate lip-speech synchronization confidence for each candidate clip. Clips with confidence scores below 3.0 are discarded. This stage filters out low-confidence audio-visual pairs after the speech and face checks.

\begin{figure*}[t]
  \centering
  \includegraphics[width=\textwidth]{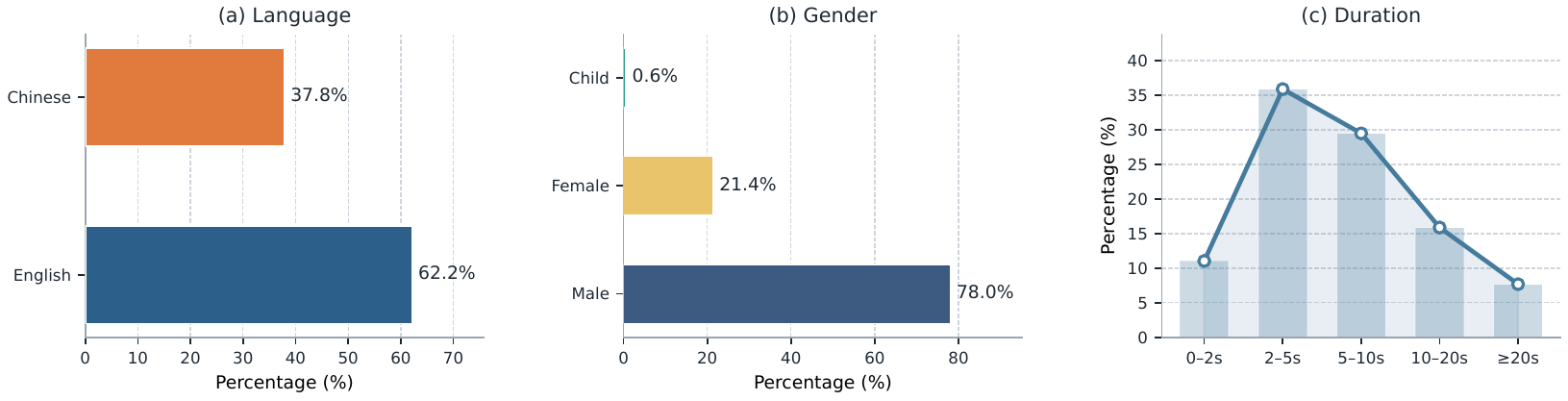}
  \caption{Corpus composition: (a)~language, (b)~gender, and (c)~clip duration.}
  \label{fig:dataset-composition}
\end{figure*}

\begin{figure*}[t]
  \centering
  \includegraphics[width=\textwidth]{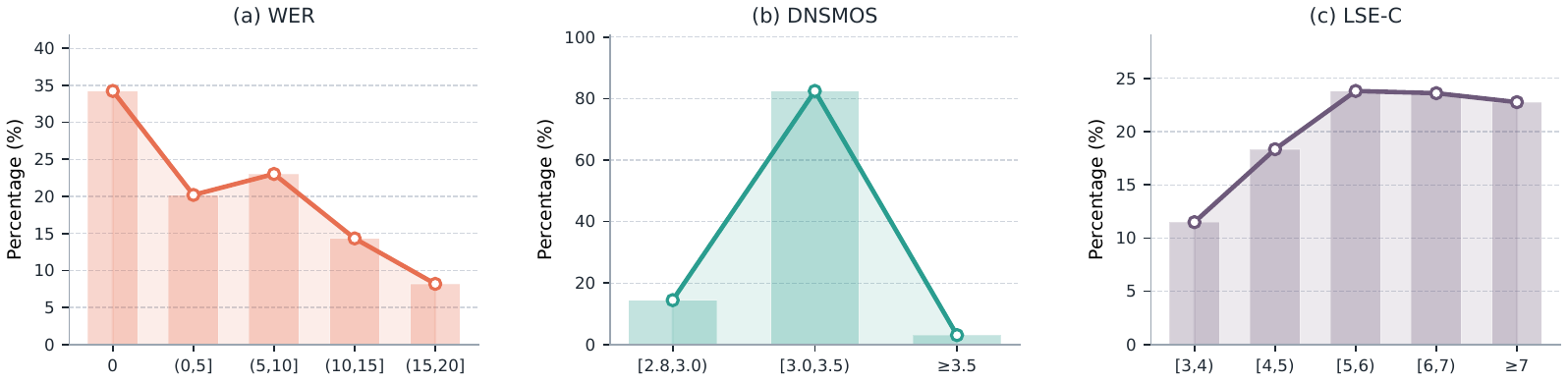}
  \caption{Quality-score distributions: (a)~inter-ASR WER, (b)~DNSMOS P.835 OVRL, and (c)~LSE-C.}
  \label{fig:dataset-quality}
\end{figure*}

\subsection{Dataset Statistics}

The curated VoxDub corpus contains $766{,}708$ clips. We characterize its composition, visual continuity, resolution, and quality-score distributions.

\subsubsection{Dataset Composition}
Fig.~\ref{fig:dataset-composition} summarizes language, speaker-category, and clip-duration distributions. We additionally examine visual continuity because speech-based boundaries can retain camera transitions within a clip. Using SceneDetect\footnote{\url{https://github.com/Breakthrough/PySceneDetect}}, we find that $97.3\%$ of clips contain a single detected scene. The median short-side resolution is $1080$\,px, and $92.8\%$ of clips have a short-side resolution above 720\,px. These statistics characterize the continuity and spatial resolution of the retained video inputs.

\subsubsection{Quality Distribution}
Fig.~\ref{fig:dataset-quality} shows the distributions of inter-ASR WER, DNSMOS P.835 OVRL, and LSE-C. Nearly 35\% of retained clips have an inter-ASR WER of zero, meaning that the two ASR hypotheses match under the applied comparison procedure. More than half have WERs below 5\%. These results indicate substantial agreement between the recognizers, which serves as a proxy for transcription reliability.

More than 80\% of clips have DNSMOS P.835 OVRL scores above 3.0, and over 70\% have LSE-C scores above 5.0. These distributions characterize the predicted audio quality and synchronization confidence of the retained corpus. They are consistent with the quality criteria used during curation, rather than independent evidence of the contribution of each filtering stage.

\section{Experiments}

\subsection{Experimental Setup}

\subsubsection{Tasks}
We evaluate VoxDub on two complementary tasks. For TTS, we use ZipVoice~\cite{zhu2025zipvoice}, a compact flow-matching zero-shot TTS model, to assess the utility of the speech--text pairs for synthesis. For AVD, we use SyncVoice~\cite{wang2025syncvoice} to assess the utility of the paired video and speech for video-conditioned generation. Together, these tasks assess the suitability of VoxDub for speech synthesis and video-conditioned dubbing.

\subsubsection{Training Data}

For TTS, we train separate Chinese and English ZipVoice models. The Chinese models are trained on VoxDub-ZH and CN-CVS for 100K steps using 4 GPUs and a per-GPU batch size of 250 seconds of speech. The English models trained on VoxDub-EN and LRS3 use 180K steps with the same GPU count and batch size. The LibriTTS model follows the official ZipVoice recipe for 60 epochs, or approximately 115K steps. We also evaluate the official ZipVoice checkpoint pretrained on the large-scale Emilia corpus as a reference model.

For AVD, we follow SyncVoice and initialize from ZipVoice weights pretrained on Emilia. We fine-tune Chinese models on VoxDub-ZH and CN-CVS and an English model on VoxDub-EN. The stated AVD training configuration is 100K steps on 2 GPUs, with a per-GPU batch size of 250 seconds of speech. For the LRS3 baseline, we directly use the official open-source checkpoint. 

\subsubsection{Evaluation Sets}

We introduce DubBench, a bilingual test set for evaluating speech synthesis and automatic video dubbing. DubBench comprises two subsets: DubBench-ZH, constructed from CMLR with 1,000 Chinese samples, and DubBench-EN, constructed from CelebV-Dub with 978 English samples. Each sample includes reference speech, target text, and the corresponding real video clip.

For TTS, we evaluate Chinese models on Seed-TTS test-zh~\cite{anastassiou2024seed} and DubBench-ZH, and English models on Seed-TTS test-en and DubBench-EN. For AVD, we use only DubBench-ZH and DubBench-EN, as this task requires paired video inputs.

\subsubsection{Evaluation Metrics}
We use objective metrics and human ratings to evaluate the generated speech.

For objective evaluation, we consider four aspects:
(1) Intelligibility is measured by WER between an ASR transcription of the generated speech and the input text. We use Whisper-large-v3~\cite{pmlr-v202-radford23a} for English and Paraformer-zh~\cite{Gao2022ParaformerFA} for Chinese. 
(2) Speaker similarity is measured by the cosine similarity between embeddings of the generated and reference speech. We extract embeddings using a WavLM-based~\cite{chen2022wavlm} ECAPA-TDNN~\cite{Desplanques2020ECAPATDNNEC} speaker verification model and denote the metric as SIM-o~\cite{le2023voicebox}.
(3) Naturalness is estimated using UT-MOS~\cite{saeki22c_interspeech}, a neural mean opinion score (MOS) predictor.
(4) Lip-speech synchronization is evaluated for AVD using LSE-C~\cite{prajwal2020lip}, computed with the pretrained SyncNet model from~\cite{li2024latentsync}. Higher LSE-C indicates greater estimated synchronization confidence between the generated speech and visible mouth motion.

For subjective evaluation, we randomly select 10 samples from each test set and collect ratings from 15 human raters. We report CMOS, SMOS, and, for AVD, Sync-MOS. CMOS evaluates naturalness relative to the reference speech on a scale from $-3$ to $3$ in increments of $1$, with higher scores indicating better perceived quality relative to the reference. SMOS evaluates speaker similarity to the reference speech on a scale from $1$ to $5$ in increments of $0.5$. Sync-MOS evaluates perceived synchronization between generated speech and visible mouth motion on the same $1$--$5$ scale with increments of $0.5$. Higher SMOS and Sync-MOS scores indicate greater perceived speaker similarity and lip-speech synchronization, respectively.

\begin{table}[t]
\centering
\caption{English TTS results on Seed-TTS test-en and DubBench-EN. WER is reported as a percentage.}
\label{tab:tts_en}
\resizebox{\linewidth}{!}{%
\begin{tabular}{lccccc|ccccc}
\toprule
\multirow{2}{*}{\textbf{Train Data}} 
& \multicolumn{5}{c|}{\textbf{Seed-TTS test-en}} 
& \multicolumn{5}{c}{\textbf{DubBench-EN}} \\
\cmidrule(lr){2-6} \cmidrule(lr){7-11}
& \textbf{WER} $\downarrow$ & \textbf{SIM-o} $\uparrow$ & \textbf{UT-MOS} $\uparrow$ & \textbf{CMOS} $\uparrow$ & \textbf{SMOS} $\uparrow$
& \textbf{WER} $\downarrow$ & \textbf{SIM-o} $\uparrow$ & \textbf{UT-MOS} $\uparrow$ & \textbf{CMOS} $\uparrow$ & \textbf{SMOS} $\uparrow$ \\
\midrule
Emilia & 1.55 & 0.696 & 3.85 & 0.19 & 4.08 & 2.40 & 0.635 & 3.58 & 0.14 & 3.86 \\
LibriTTS & 3.92 & 0.603 & 3.69 & 0.07 & 3.79 & 8.38 & 0.511 & 3.46 & 0.03 & 3.51 \\
\midrule
LRS3 & 12.54 & 0.448 & 3.36 & -0.16 & 3.47 & 12.34 & 0.506 & 3.06 & -0.07 & 3.55 \\
VoxDub-EN & 3.01 & 0.601 & 3.92 & 0.14 & 3.96 & 4.34 & 0.562 & 3.42 & 0.09 & 3.67 \\
\bottomrule
\end{tabular}%
}
\end{table}

\begin{table}[t]
\centering
\caption{Chinese TTS results on Seed-TTS test-zh and DubBench-ZH. WER is reported as a percentage.}
\label{tab:tts_zh}
\resizebox{\linewidth}{!}{%
\begin{tabular}{lccccc|ccccc}
\toprule
\multirow{2}{*}{\textbf{Train Data}} 
& \multicolumn{5}{c|}{\textbf{Seed-TTS test-zh}} 
& \multicolumn{5}{c}{\textbf{DubBench-ZH}} \\
\cmidrule(lr){2-6} \cmidrule(lr){7-11}
& \textbf{WER} $\downarrow$ & \textbf{SIM-o} $\uparrow$ & \textbf{UT-MOS} $\uparrow$ & \textbf{CMOS} $\uparrow$ & \textbf{SMOS} $\uparrow$
& \textbf{WER} $\downarrow$ & \textbf{SIM-o} $\uparrow$ & \textbf{UT-MOS} $\uparrow$ & \textbf{CMOS} $\uparrow$ & \textbf{SMOS} $\uparrow$ \\
\midrule
Emilia & 1.42 & 0.750 & 3.14 & 0.06 & 4.03 & 1.72 & 0.740 & 3.55 & -0.03 & 4.11 \\
\midrule
CN-CVS & 12.52 & 0.482 & 2.81 & -0.22 & 3.64 & 7.17 & 0.593 & 2.71 & -0.25 & 3.48 \\
VoxDub-ZH & 2.63 & 0.613 & 3.22 & -0.02 & 3.88 & 3.08 & 0.652 & 2.70 & -0.18 & 3.97 \\
\bottomrule
\end{tabular}%
}
\end{table}

\subsection{TTS Results}

The TTS experiments assess the utility of VoxDub's speech--text pairs on the Seed-TTS and DubBench test sets. English comparisons include LibriTTS, LRS3, and the Emilia reference model; Chinese comparisons include CN-CVS and the Emilia reference model.

Table~\ref{tab:tts_en} reports the English results. The Emilia reference model achieves the lowest WER and highest SIM-o and SMOS on both test sets, although it does not obtain the highest UT-MOS on Seed-TTS test-en. Relative to Emilia, the LibriTTS model has higher WER and lower speaker similarity, particularly on DubBench-EN, where its WER is 8.38. The difference is consistent with the potential benefit of large-scale TTS-oriented data.

Among the compared audio-visual corpora, VoxDub-EN improves on LRS3 across all reported metrics on both English test sets. WER decreases from 12.54 to 3.01 on Seed-TTS test-en and from 12.34 to 4.34 on DubBench-EN. LRS3 has the lowest SIM-o and UT-MOS on both sets; its SMOS is the lowest on Seed-TTS test-en, but exceeds that of LibriTTS on DubBench-EN. Relative to LibriTTS, VoxDub-EN achieves lower WER and higher CMOS and SMOS on both sets. Its SIM-o is slightly lower on Seed-TTS test-en and higher on DubBench-EN, while its UT-MOS is higher on the former and lower on the latter. These results support the utility of VoxDub-EN for TTS under the reported settings. 

Table~\ref{tab:tts_zh} reports the Chinese results. Compared with CN-CVS, VoxDub-ZH reduces WER from 12.52 to 2.63 on Seed-TTS test-zh and from 7.17 to 3.08 on DubBench-ZH, while achieving higher SIM-o, CMOS, and SMOS on both sets. UT-MOS is higher on Seed-TTS test-zh and nearly unchanged, but numerically lower, on DubBench-ZH (2.70 versus 2.71). The Emilia reference model retains lower WER and higher speaker similarity on both sets; VoxDub-ZH has higher UT-MOS on Seed-TTS test-zh but lower UT-MOS on DubBench-ZH. Overall, VoxDub improves intelligibility and speaker similarity relative to the compared audio-visual corpora in these TTS experiments. These results support the use of VoxDub's curated speech--text pairs for intelligible speech synthesis with consistent speaker characteristics.

\subsection{Audio Quality Comparison}

\begin{figure}[t]
\centering
\includegraphics[width=0.75\linewidth]{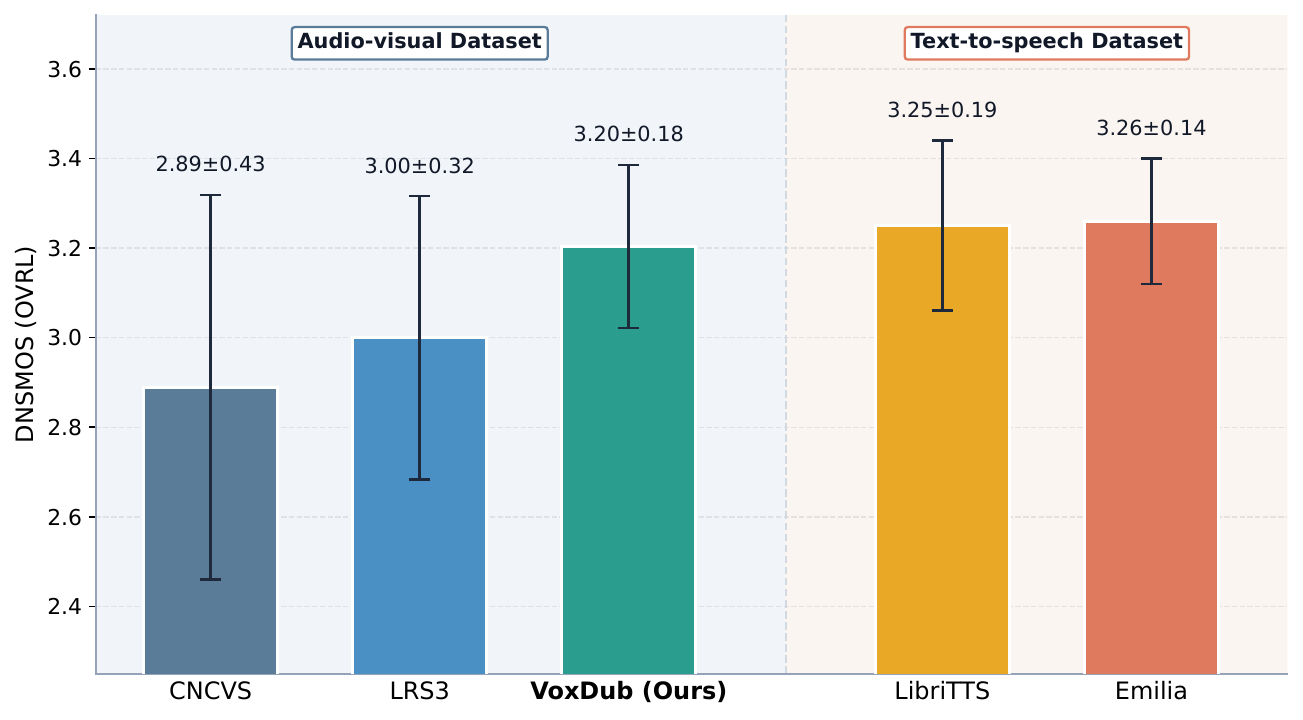}
\caption{Comparison of DNSMOS OVRL scores across different corpora. Error bars denote standard deviations.}
\label{fig:dnsmos_ovr_compare}
\end{figure}
To complement the downstream evaluation, Fig.~\ref{fig:dnsmos_ovr_compare} compares mean DNSMOS OVRL scores and their standard deviations across corpora. DNSMOS is a reference-free predictor of perceived speech quality. We randomly sample 100K utterances from CN-CVS, LRS3, and VoxDub to compute these statistics; the Emilia and LibriTTS values are taken from~\cite{he2024emilia}.

In Fig.~\ref{fig:dnsmos_ovr_compare}, VoxDub has a higher mean score than CN-CVS and LRS3 and a mean close to those reported for Emilia and LibriTTS. This ordering is consistent with its favorable TTS results relative to the compared audio-visual corpora. Together, the DNSMOS scores and TTS results support the suitability of VoxDub's speech recordings for speech synthesis training.

\begin{table}[t]
\centering
\caption{AVD results on DubBench-EN and DubBench-ZH. WER is reported as a percentage.}
\label{tab:avd}
\resizebox{\linewidth}{!}{%
\begin{tabular}{llccccccc}
\toprule
\textbf{Lang} & \textbf{Train Data}  & \textbf{WER} $\downarrow$ & \textbf{SIM-o} $\uparrow$ & \textbf{UT-MOS} $\uparrow$ & \textbf{LSE-C} $\uparrow$ &  \textbf{CMOS} $\uparrow$ & \textbf{SMOS} $\uparrow$ & \textbf{Sync-MOS} $\uparrow$ \\
\midrule
\multirow{2}{*}{EN} & LRS3 & 6.10 & 0.652 & 3.51 & 7.10 & -0.06 & 3.75 & 4.27 \\
& VoxDub-EN & 5.44 & 0.649 & 3.54 & 6.29 & -0.02 & 3.83 & 4.12 \\
\midrule
\multirow{2}{*}{ZH} & CN-CVS & 1.40 & 0.728 & 2.78 & 6.05 & 0.06 & 3.65 & 4.01 \\
& VoxDub-ZH & 1.41 & 0.730 & 2.74 & 5.79 & 0.09 & 3.71 & 4.07 \\
\bottomrule
\end{tabular}%
}
\end{table}

\subsection{AVD Results}

The AVD experiments evaluate VoxDub as a fine-tuning corpus for video-conditioned speech generation. Following SyncVoice, the models adapt a pretrained ZipVoice speech generator using video-conditioned modules.

Table~\ref{tab:avd} shows broadly similar speech-side results across the compared corpora. On DubBench-EN, VoxDub-EN yields lower WER and slightly higher UT-MOS than LRS3, with slightly lower SIM-o. On DubBench-ZH, VoxDub-ZH has nearly the same WER as CN-CVS, slightly higher SIM-o, and slightly lower UT-MOS. The VoxDub models obtain higher CMOS and SMOS point estimates in both languages. The smaller speech-side differences observed in AVD may be partly explained by the pretrained TTS model, which provides a shared foundation for intelligibility, speaker similarity, and naturalness before adaptation to video-conditioned dubbing.

Synchronization results are mixed. VoxDub has lower LSE-C than LRS3 on DubBench-EN (6.29 versus 7.10) and lower LSE-C than CN-CVS on DubBench-ZH (5.79 versus 6.05). The subjective Sync-MOS scores are close: VoxDub is lower on English (4.12 versus 4.27) and higher on Chinese (4.07 versus 4.01). These point estimates support broadly comparable perceived synchronization in the evaluated samples, without establishing equivalence or a synchronization advantage. Taken together, the results support VoxDub as a usable AVD fine-tuning corpus with competitive speech quality and subjective synchronization under the reported conditions.

\section{Conclusion}

We presented VoxDub, a bilingual audio-visual dataset for automatic video dubbing. Its curation pipeline combines speech-centered segmentation with checks on transcription agreement, audio quality, face visibility, and lip-speech synchronization. The resulting corpus supports both speech--text and video--speech modeling.

In the reported TTS experiments, VoxDub improves intelligibility and speaker similarity relative to the compared audio-visual corpora. In AVD fine-tuning, it yields competitive speech-side results and broadly comparable subjective synchronization. These findings support its utility for the evaluated speech generation tasks. VoxDub also offers a potential data resource for broader multimodal speech research, including audio-visual speech representation learning, lip reading, and talking-face speech generation.


%
%

\bibliographystyle{splncs04}
\bibliography{mybibliography}



\end{document}